\documentstyle[12pt]{article}
\def\beqa{\begin{eqnarray}}
\def\eeqa{\end{eqnarray}}
\def\beq{\begin{equation}}
\def\eeq{\end{equation}}

\def\half{\frac{1}{2}}

\def\al{{\alpha}}
\def\be{{\beta}}

\def\del{{\delta}}

\def\sig{{\sigma}}

\def\l{{\cal L}}

\def \phan{\phantom}

\def\ie{{\it i.e. }}
\def\eg{{\it e.g. }}

\def\pr{{\it Phys. Rev.}\ }
\def\prl{{\it Phys. Rev. Lett.}\ }
\def\pl{{\it Phys. Lett.}\ }

\def\cqg{{\it Class. Quantum Grav.}\ }

\def\grg{{\it Gen. Relativ. Grav.}\ }

\def\ncim{{\it Nuovo Cim.}\ }

\def\rmp{{\it Rev. Mod. Phys.}\ }

\def\ass{{\it Astr. and Space Sci.}\ }
\begin{document}
\def\bib#1{[{\ref{#1}}]}

\title{Geometric classification of the torsion tensor of space-time}

 \author{{S. Capozziello$^{1,3}$ G. Lambiase $^{1,3}$ and C. Stornaiolo$^{2,3}$}
\\
{\em $~^{1}$ Dipartimento di Fisica``E. R. Caianiello'',}
\\ {\em Universit\`{a} di Salerno, I-84081 Baronissi, Salerno,
Italy.} \\ {\em $~^{2}$ Dipartimento di Scienze Fisiche,
Universit\`{a} ``Federico II'' di Napoli,}\\ {\em Complesso
Universitario di Monte S. Angelo,}\\ {\em Via Cintia, Edificio
N', I-80126 Napoli, Italy }
\\ {\em $~^{3}$ Istituto Nazionale di Fisica Nucleare, Sezione di
Napoli,}
\\ {\em Complesso Universitario di Monte S. Angelo}\\ {\em Via Cintia, Edificio N' I-80126 Napoli, Italy.}}
          \date{\today}
          \maketitle

\vspace{5. mm}
e-mail addresses: \\
 capozziello@sa.infn.it \\
 lambiase@sa.infn.it\\
 cosmo@na.infn.it \\

\vspace{3. mm}

  \noindent{\footnotesize We wish to dedicate this paper to the
memory of Ruggiero de Ritis who passed away while this paper was
being completed. S.C. and C.S. mourn the loss of their mentor and
lament the loss of a dear friend. }

\begin{abstract}
Torsion appears in literature in quite different forms.
Generally, spin is  considered to be  the source of torsion, but
there are several other possibilities in which torsion emerges in
different contexts.  In some cases  a phenomenological
counterpart is absent, in some other cases torsion  arises from
sources without spin as a gradient of a scalar field. Accordingly,
we propose  two classification schemes.  The {\em first} one is
based on the possibility to construct torsion tensors from the
product of a covariant bivector and a vector and their respective
space-time properties. The {\em second} one is obtained by
starting from the decomposition of torsion into three irreducible
pieces. Their space-time properties again lead to a complete
classification. The classifications found are given in a {\bf
U}$_4$, a four dimensional space-time where the torsion tensors
have some peculiar properties. The irreducible decomposition is
useful since  most of the phenomenological work done for torsion
concerns four dimensional cosmological models. In the second part
of the paper two applications of these classification schemes are
given. The modifications of energy-momentum tensors  are
considered that arise due to different sources of torsion.
Furthermore, we analyze   the contributions of torsion to shear,
vorticity, expansion and acceleration. Finally the generalized
Raychaudhuri equation is discussed.
\end{abstract}

\section{ Introduction}
The issue to enlarge the classical scheme of General Relativity is
felt strongly today since several questions strictly depend on the
fact if the spacetime connection is symmetric or not. General
Relativity is essentially a classical theory which does not take into
account quantum effects. These ones must be considered in any theory
which deals with gravity on a fundamental level.  Passing from {\bf
  V}$_4$ to {\bf U}$_4$ manifolds, is the first straightforward
generalization which tries to include the spin fields of matter into
the same geometrical scheme of General Relativity. The
Einstein--Cartan--Sciama--Kibble (ECSK) theory is one of the most
serious attempts in this direction \bib{hehl}. However, this mere
inclusion of spin matter fields does not exhaust the role of torsion
which seems to give important contributions in any fundamental theory.

For example, a torsion field appears in (super)string theory if we
consider the fundamental string modes; we need, at least, a scalar
mode and two tensor modes: a symmetric and an antisymmetric one. The
latter one, in the low energy limit for the effective string action, gives
the effects of a torsion field \bib{GSW}.

Furthermore, several attempts of unification between gravity and
electromagnetism have to take into account torsion in four and in
higher--dimensional theories such as Kaluza-Klein ones \bib{GER}.

Any theory of gravity considering twistors needs the inclusion of
torsion \bib{HOW} while supergravity is the natural arena in which
torsion, curvature and matter fields are treated in an analogous way
\bib{LOS}.

Besides, several people agree with the line of thinking that torsion
could have played some specific role in the dynamics of the early
universe and, by the way, it could have yielded macroscopically
observable effects today. In fact, the presence of torsion naturally
gives repulsive contributions to the energy-momentum tensor so that
cosmological models become singularity-free \bib{SIV}. This feature,
essentially, depends on spin alignments of primordial particles which
can be considered as the source of torsion \bib{MAR}. If the universe
undergoes one or several phase transitions, torsion could give rise to
topological defects (\eg torsion walls \bib{FIG}) which today can act
as intrinsic angular momenta for cosmic structures as galaxies.
Furthermore, the presence of torsion in an effective energy-momentum
tensor alters the spectrum of cosmological perturbations giving
characteristic lengths for large scale structures \bib{ZZI}.

All these arguments, and several more, do not allow to neglect torsion
in any comprehensive theory of gravity which takes into account the
non gravitational counterpart of the fundamental interactions.

However, in most articles, a clear distinction is not made among the
different kinds of torsion. Usually torsion is related with the spin
density of matter, but very often there are examples where it cannot
be derived from it and assumes meanings quite different from the
models with spinning fluids and particles. It can be shown that there
are many independent torsion tensors with different properties.

In order to clarify these points, we have found two classification
schemes of torsion tensors based on the geometrical properties of
vectors and bivectors that can be used to decompose them. The
first classification found is based on the possibility to
construct  the torsion tensors as the tensor product of a simple
covariant bivector and a contravariant vector. Such objects are
well understood in General Relativity and they can be easily
classified
\bib{mtw}. Then classifying  all the possible
combinations, we find 24 independent tensors. We call these
tensors {\em elementary torsions}.

The second classification follows from the decomposition at one point
of a {\bf U}$_4$ space-time of the torsion tensors into three
irreducible tensors with respect to the Lorentz group. Again we could
use vectors and bivectors to identify their geometrical properties. It
follows that the elements of the second classification are generally
expressed as a combination of ``elementary torsion tensors'', while
the ``elementary torsion tensors'' are generally non-irreducible.

One of the main results of our classifications is that in many
theories, such as the Einstein-Cartan-Sciama-Kibble theory, torsion is
related to its sources by an algebraic equation; it follows that
these two classifications clarify the nature of the sources too.
This feature leads to recognize which tensors can be generated by
the spin and which not and which do not even have a physical
origin.

To our knowledge, two other classifications have been already
proposed. The first one was given in \bib{baker} and it is based on
the properties of the Riemann and Ricci tensors as defined in a {\bf
  U}$_4$ spacetime compared with the Weyl and Ricci tensors as defined
in a {\bf V}$_4$ spacetime. The second, given in \bib{OTM}, deals with
the algebraic classification of spacetimes with torsion following from
the application of the BRST operator.

The classifications of the torsion tensors show of how the different
sources of torsion can influence the physical phenomena. It is well
known\bib{hehl} that the ECSK theory can be recast in a nonminimal
coupling version of ordinary General Relativity, where the
energy-momentum tensor is modified by the torsion sources. This
contributions are calculated and discussed for all types found in the
classification.

~From a phenomenological point of view, the torsion theories may
be relevant in cosmology. This because the kinematical
quantities, shear, vorticity, acceleration, expansion and their
evolution equations are modified by the presence of torsion.

The paper is organized as follows. In Sec.\ref{gener},
we give general notations and definitions following
\bib{hehl} and references therein.

In Sec.\ref{element}, we give the first classification
through the elementary torsion tensors and in \ref{irreduc}
we specify the elements of the second classification. In
Sec.\ref{examples}, without any pretence of completeness,
we review some  relevant  torsion models that have
been used in literature. As a result it is shown that just
three of them can be related to the usual spin sources
generally treated in literature \bib{hehl}; while the
other kinds have sources of a physically  different origin
or even, to our knowledge, their possible sources have not
a clear physical interpretation.

In  Secs.\ref{emtens} and \ref{svea}, some applications are
given. In the former, the contributions to the energy-momentum
tensor in the ECSK theory are calculated. In the latter, the
contributions of torsion to the different components of the
gradient of the $4$-velocity $U_a$ are obtained. This discussion
is completed finding the most general expression of the
Raychaudhuri equation. General discussion and conclusions are
given in Sec.\ref{conclusions}.

\section{ General remarks}\label{gener}
In this section, we give general definitions of torsion and
associated quantities which, below, will be specified in the
particular {\bf U}$_{4}$ spacetimes. We shall use, essentially,
the notation in \bib{hehl}.

\subsection{\normalsize\bf Definitions}

The torsion tensor $S_{ab}^{\phan{ab}c}$ it the antisymmetric part
of the affine connection coefficients $\Gamma_{ab}^{c}$, that is
 \beq
 \label{t1}
S_{ab}^{\phan{ab}c}=\frac{1}{2}\left(\Gamma_{ab}^{c}-\Gamma_{ba}^{c}\right)
\equiv\Gamma_{[ab]}^{c}\,, \eeq
where $a,b,c = 0,\dots 3$.

In General Relativity it is postulated that
$S_{ab}^{\phan{ab}c}=0$. It is a general convention to call {\bf
U}$_4$ a $4$-dimensional space-time manifold endowed with metric
and torsion. The manifolds with metric and without torsion are
labeled as {\bf V}$_4$ (see
\bib{schouten}).

Often in the calculations, torsion occurs in linear combinations as
in the {\it contortion tensor}, defined as
 \beq
\label{t3}
K_{ab}^{\phantom{ab}c}=-S_{ab}^{\phantom{ab}c}-S^{c}_{\phan{c}a b}
+
S^{\phantom{a}c}_{b\phantom{c}{a}}=-K^{\phantom{a}c}_{a\phan{c}b}\,,
\eeq

and in the {\it modified torsion tensor}

\beq \label{t4} T_{ab}^{\phantom{ab}c}=
S_{ab}^{\phantom{ab}c}+2\delta^{\phan{[a}c}_{[a}S_{b]}
 \eeq
 where $S_a\equiv S_{ab}^{\phantom{ab}b}$.

According to these definitions, it follows that the affine
connection can be written as \beq \label{t5}
\Gamma_{ab}^{c}=\left\{^{c}_{ab}\right\}-K_{ab}^{\phantom{ab}c}\,,
\eeq where $\left\{^{c}_{ab}\right\}$ are the usual
Christoffel symbols of the symmetric connection. The
presence of torsion in the affine connection implies that
the covariant derivatives of a scalar field $\phi$ do not
commute, that is
\begin{equation}\label{scalarder}
 \tilde \nabla_{[a}\tilde\nabla_{b]}\phi=-S_{ab}^{\phan{ab}c}\tilde\nabla_{c}\phi;
\end{equation}
and for a vector $v^a$ and a covector $w_a$, one has the following
relations

\begin{equation}\label{doppiader1}
 (\tilde\nabla_{a}\tilde\nabla_{b} - \tilde\nabla_{b}\tilde\nabla_{a})v^{c}=
  R_{abd}^{\phantom{abd}c}v^d
 -2S_{ab}^{\phan{ab}d}\tilde\nabla_{d}v^c,
 \end{equation}
 and

\begin{equation}\label{doppiader2}
 (\tilde\nabla_{a}\tilde\nabla_{b} - \tilde\nabla_{b}\tilde\nabla_{a})w_{d} =
  R_{abc}^{\phantom{abc}d}w_{d}
 -2S_{ab}^{\phan{ab}d}\tilde\nabla_{d}w_{c}
\end{equation}

where the Riemann tensor is defined as
\begin{equation}\label{riemann}
 R_{abc}^{\phantom{abc}d}=\partial_{a}\Gamma_{bc}^{d}-\partial_{b}\Gamma_{ac}^{d}
 +\Gamma_{ae}^{d}\Gamma_{bc}^{e} - \Gamma_{be}^{d}\Gamma_{ac}^{e}.
\end{equation}

The contribution to the Riemann tensor of torsion can be explicitly given by
\begin{equation}\label{riexpanded}
R_{abc}^{\phantom{abd}d} =R_{abc}^{\phantom{abd}d}(\{\}) -
 \nabla_{a}K_{bc}^{\phantom{b]c}d} +  \nabla_{b}K_{ac}^{\phantom{ac}d}
+ K_{ae}^{\phantom{ae}d}K_{bc}^{\phantom{bc}e}-
K_{be}^{\phantom{be}d}K_{ac}^{\phantom{ac}e}
\end{equation}
where $R_{abc}^{\phantom{abc}d}(\{\})$ is the tensor generated by
the Christoffel symbols. The symbols $\tilde\nabla$ and $\nabla$
have been used to indicate the covariant derivative with and
without torsion respectively.

~From Eq.(\ref{riexpanded}), the expressions for the Ricci tensor
and the curvature scalar are
\begin{equation}\label{ricci}
 R_{ab}= R_{ab}(\{\}) - 2\nabla_{a}S_{c} + \nabla_{b}K_{ac}^{\phantom{ac}b}
+ K_{ae}^{\phantom{ae}b}K_{bc}^{\phantom{bc}e}-
2S_eK_{ac}^{\phantom{ac}e}
\end{equation}
and
\begin{equation}\label{curvscalar}
 R=R(\{\}) - 4 \nabla_{a}S^{a} + K_{ceb}K^{bce} - 4 S_aS^a.
\end{equation}

\subsection{\normalsize\bf Tetrads and bivectors }\label{bivectors}

In the following, we will use the tetrad fields. They are defined
at each point of the manifold as a base of orthonormal vectors
$e_{A}^{a}$, where $A,B,C\dots=0,1,2,3$ label the tetrad vectors
and $a,b,c,\dots=0,1,2,3$ are the component indices; $e_{0}^{a}$
is a time-like vector and $e_{I}^{a}$ is space-like.

Correspondingly, a cotetrad $e_{a}^{A}$ is defined such that
\begin{equation}\label{delta1}
  e_{A}^{a}e_{b}^{A}=\delta^{a}_{b}\,,
  \end{equation}
\begin{equation}\label{delta2}
 e_{A}^{a}e_{a}^{B}=\delta^{B}_{A}\,.
\end{equation}

The tetrad metric is

\begin{equation}\label{metrad}
\eta_{AB}=\eta^{AB}= diag(-1,1,1,1),
\end{equation}
then the space-time metric can be reconstructed  in the
following way

\begin{equation}\label{metrica}
g_{ab} =\eta_{AB}e_{a}^{A}e_{b}^{B}.
\end{equation}
In the construction of the torsion tensors, it will be useful to
consider expression of the simple bivectors. These are given by
the skew-symmetric tensor product of two vectors. Generally a
 bivector $B^{ab}$ is simple, if and only if it satisfies the equation

\begin{equation}\label{simbiv}
  B^{[ab}B^{c]d}=0.
\end{equation}
By the tetrad vectors in a $N$-dimensional manifold, one can
construct the $N(N-1)/2$ simple bivectors
\begin{equation}\label{simple}
  F_{AB}^{ab}= e_{A}^{[a}e_{B}^{b]}
\end{equation}

and  any bivector $B^{ab}$ is expressed as
\begin{equation}\label{biv} B^{ab}=
B^{AB}e_{A}^{a}e_{B}^{b}
\end{equation}
with $B^{AB}=-B^{BA}$.

\subsection{\normalsize\bf  Decomposition of  torsion
and its properties in  U$_4$}\label{gendec}

An important property of torsion is that it can be decomposed with
respect to the Lorentz group into three irreducible tensors,
  i.e. it can be written as
\begin{equation}\label{decomposition}
 S_{ab}^{\phantom{ab}c}= {}^T S_{ab}^{\phantom{ab}c} +
{}^A S_{ab}^{\phantom{ab}c}+ {}^V S_{ab}^{\phantom{ab}c}.
\end{equation}
Torsion has $24$ components, of which ${}^T S_{ab}$ has $16$
components, ${}^A S_{ab}$ has $4$ and ${}^V S_{ab}$ has the
remaining $4$ components
\bib{mccrea},\bib{tsamparlis},\bib{affine}.

 One has
\begin{equation}\label{vector}
 {}^V S_{ab}^{\phantom{ab}c}=\frac{1}{3}(S_{a}\delta_{b}^{c}-
 S_{b}\delta_{a}^{c}),
\end{equation}
where $S_a =  S_{ab}^{\phantom{ab}b}$,
\begin{equation}\label{axial}
 {}^A S_{ab}^{\phantom{ab}c}=g^{cd}S_{[abd]}
\end{equation}
which is called the axial (or totally anti-symmetric) torsion, and
\begin{equation}\label{tensor}
 {}^T S_{ab}^{\phantom{ab}c}= S_{ab}^{\phantom{ab}c} - {}^A
 S_{ab}^{\phantom{ab}c}- {}^V S_{ab}^{\phantom{ab}c}
\end{equation}
which is the traceless non totally anti-symmetric part of torsion.

For the sake of brevity, in the following, we will refer respectively to
the tensor (\ref{vector}) as a V-torsion, to  the tensor
(\ref{axial}) as an A-torsion and  to the tensor (\ref{tensor}) as a
T-torsion.

The dual operation (see
\bib{mccrea},\bib{affine}) defined as
\begin{equation}\label{dual}
{}^{+}
S_{ab}^{\phantom{ab}c}=\frac{1}{2}\epsilon^{de}_{\phantom{de}ab}
S_{de}^{\phantom{de}c}
\end{equation}
has the relevant property, that it associates an A-torsion tensor
to a V-torsion tensor
and vice versa. Then it associates a T-torsion to a
T-torsion.
\subsection{\normalsize\bf The Einstein-Cartan
equations}\label{ecks}

The introduction of  torsion as an extension of the gravitational
field theories has some relevant consequences.

The closest theory to General Relativity is the
Einstein-Cartan-Sciama-Kibble (ECSK) theory. It is described by

\beq \label{t6}
 L=\sqrt{-g}\left(\frac{R}{2k}+{\l}_{m}\right),
\eeq which is  the lagrangian density of General Relativity
depending on the metric tensor $g_{ab}$ and on the connection
$\Gamma_{ab}^{c}$, where $R$ is the curvature scalar
(\ref{curvscalar}) and ${\l}_{m}$ the Lagrangian function of
matter fields, which yields

 \beq \label{t7}
\mbox{{\bf t}}^{ab}=\frac{\delta {\cal
L}_{m}}{\delta g_{ab}},
 \eeq
which is the symmetric stress--energy tensor while
 \beq \label{t8}
\tau_{c}^{\phan{c}ba}=\frac{\delta {\cal
L}_{m}}{\delta K_{ab}^{\phan{ab}c}},
 \eeq
is the source of torsion. In many instances, it can be identified
with a spin density. But, as will be clear from the following
sections, there are many cases in which the source of the torsion
field (\ref{t8}) is not spin.

~From the variation of (\ref{t6}) and introducing the canonical
energy-momentum tensor
\begin{equation}\label{canonical}
{\bf \Sigma^{ab}}= {\bf t^{ab}} + {}^*\tilde\nabla_{c}(\tau^{abc}
- \tau^{bca} +\tau^{cab})\, ,
\end{equation}
where we used the abridged notation
${}^*\tilde\nabla_{c}:=\tilde\nabla_{c} +2S_{cd}^{\phantom{cd}d}$,
the following field equations are derived \bib{hehl}
\begin{equation}\label{t9}
   G^{ab}=k  {\bf\Sigma}^{ab}\, ,
\end{equation}
and
 \beq \label{t10}
 T_{ab}^{\phan{ab}c}=k\tau_{ab}^{\phan{ab}c}\,,
\eeq
 where
$k=8\pi G,\,\,c=1$.

Equation (\ref{t9}) is the generalizes the Einstein equations in a
$U_4$.

Unlike Eq.(\ref{t9}), Eq.(\ref{t10}) is algebraic so that it is
always possible to cast Eq.(\ref{t9}) in a pure Einstein one, by
substituting the torsion terms with their sources. It results in
defining an effective energy--momentum tensor as the source of
the Riemannian part of the Einstein tensor \bib{hehl}. In doing
so, one obtains

\beq \label{t11}
  G^{ab}(\{\})=k\tilde{\mbox{{\bf t}}}^{ab},
  \eeq

where $G^{ab}(\{\})$ is the Riemannian part of the Einstein
tensor. The effective energy--momentum tensor is

 \beqa
\label{t12}
\tilde{\mbox{{\bf t}}}^{ab}&=&\mbox{{\bf t}}^{ab}+
k\left[-4\tau^{ac}_{\phantom{ac}[d} \tau^{bd}_{\phantom{bd}c]}-
2\tau^{acd}\tau^{b}_{cd}+\tau^{cda}\tau_{cd}^{\phantom{cd}b}\right.\nonumber\\
&+&\left.\half
g^{ab}(4\tau^{\phantom{e}c}_{e\phantom{c}[d}\tau^{ed}_{\phantom{ed}c]}
+\tau^{ecd}\tau_{ecd})\right].
 \eeqa

The tensor ${\bf t}^{ab}$ can be of the form

 \beq \label{t13} \mbox{{\bf
t}}^{ab}=(p+\rho)u^{a}u^{b}-pg^{ab} ,
\eeq

if standard perfect--fluid matter is considered. But when spin
fluids are considered, one has to define a different
stress--energy tensor in which the spin contributions are taken
into account as in
\bib{deritis},\bib{drls},\bib{ray}, \bib{obukhov}.

\section{  Elementary torsion tensors: a first
classification}\label{element}

It can be observed that a tensor with all  properties of torsion
can be constructed as the tensor product of a bivector $F_{ab}$
with a vector $\Sigma^{c}$.

It is well known that any generic bivector in a four dimensional
manifold can be always reduced into the sum of two simple
bivectors with a particular choice of the coordinates (see e.g.
\bib{mtw}). In analogy with the electromagnetic case, we can  call
the bivector with the timelike vector, the {\it electric} term and
the one with two spacelike vectors, the {\it magnetic} term and
label them respectively with $E_{ab}$ and $B_{ab}$.
 Then we can
introduce the concept of elementary torsion tensor given as the
tensor product of a {\it simple} bivector with a vector.

 We
say that a bivector $A_{ab}$ and a vector $V^{c}$ are orthogonal
if $V^aA_{ab}=0$.

 We  consider only the cases where any four-vector $\Sigma^c$
 is either orthogonal  to a simple bivector or  is one of its
components. All the other possible cases are combinations of these
two cases.

Then the $24$ elementary torsion tensors can be classified
according to the space-time properties of their bivectors and the
corresponding vectors.

At this point an important remark is necessary. Any generic
torsion tensor can be decomposed in terms of these elementary
parts. Let us practically construct the elementary torsion tensors
by the vectors of a tetrad. In general, we have
\begin{equation}\label{elementa}
S_{ABCab}^{(el) c}=e^{A}_{[a}e^{B}_{b]}e_{C}^{c}\,,
\end{equation}
and then any torsion tensor can be expressed as
\begin{equation}\label{elementa1}
S_{ab}^{\phan{ab}c}=
S_{AB}^{\phan{AB}C}e^{A}_{[a}e^{B}_{b]}e_{C}^{c}\,,
\end{equation}
where the coefficients have to be
\begin{equation}\label{elementa2}
S_{AB}^{\phan{AB}C}=S_{ab}^{\phan{ab}c}
e_{A}^{[a}e_{B}^{b]}e^{C}_{c}\,.
\end{equation}

The classification of elementary torsion tensors in which
$\Sigma^{a}$ does not lie on the plane defined by the
bivector is then the following:

a) if $E_{ab}$ is a bivector obtained from the antisymmetric
tensor product of a timelike covector and a spacelike covector,
$\Sigma^a$ must be any spacelike vector  orthogonal to $E_{ab}$.
The pure electric case is represented just by one family of
tensors. It will be labeled with the symbol $ Es$;

b) in the pure magnetic case, one has that $\Sigma^a$ can be
either a spacelike vector, a timelike vector or a null vector,
leading to three family of tensors. These three families will be
labeled respectively as $ Bs$, $ Bt$ and $Bn$;

c) In the null case, it turns out that there are two
possibilities for $\Sigma^a$, \ie it can be either a null
vector or a spacelike vector. The labels will be $Nn$ and $
Ns$ respectively.

Regarding the case in which the vector $\Sigma^{a}$ lies on the
plane described by the bivector, it can be noted that, if $B\equiv
C$ in (\ref{elementa}), we have
 V-torsions.

Finally, let us note that
 the previous discussion changes if a  null tetrad, defined
by $l^a = e_{0}^{a}-e_{1}^{a}$, $n^a = e_{0}^{a}+e_{1}^{a}$,
$m^{a}=e_{2}^{a}-ie_{3}^{a}$ and ${m^*}^{a}=e_{2}^{a}+ie_{3}^{a}$,
is considered.

In this case,  it follows that the elementary torsions like
\begin{equation}\label{elementt}
S_{ab}^{\phan{ab}c}=m_{[a}l_{b]}l^{c}
\end{equation}
bear all properties of a T-torsion.

\section{ Classification in terms of irreducible tensors in four
dimensions}\label{irreduc}

To classify the torsion tensors, according to their irreducible
properties, let us first consider  the V-torsion. It follows,
from  Eq.(\ref{vector}), that the V-torsion is characterized by a
covector
\begin{equation}\label{vector2}
  S_a= S_{ab}^{\phan{ab}b}.
\end{equation}
$S_a$ can be either time-like, space-like or light-like. So we
have three different possible types of V-torsion, which can be
labeled respectively by  the symbols $Vt$, $Vs$ and $V\ell$.

It can be noted that the V-torsion is expressed as a combination
of elementary torsion tensors.

~From Eqs.(\ref{vector}) and (\ref{delta1}), it follows that
\begin{equation}\label{vector3}
{}^V S_{ab}^{\phantom{ab}c}=\frac{2}{3}S_{[a}e^{A}_{b]}e_{A}^{c}
\end{equation}

The A-torsion can be expressed by the equation

\begin{equation}\label{hodge}
  {}^A S_{abc}= \epsilon_{abcd}\sigma^d.
\end{equation}
Its properties can be characterized by the space-time properties
of the vector $\sigma^d$. As for  the V-torsion, we label the
A-torsion with $At$, $As$ or $A\ell$ depending on whether the
vector $\sigma^a$ is time-like, space-like or light-like.

The statement given in \S\ref{gendec} can be proved here by direct
calculation. In fact, we have
\begin{equation}\label{vtoa}
  \epsilon^{de}_{\phantom{de}a b}S_{[d}\delta_{e]}^{c}=
  \epsilon^{dc}_{\phantom{de}a b}S_d\,,
\end{equation}
which is an A-torsion, on the other hand
\begin{equation}\label{atov}
  \epsilon^{ef}_{\phantom{de}a b}
  \epsilon_{def}^{\phan{def}c}S^d = S_{[a}\delta_{b]}^{c}\,,
\end{equation}
which is a V-torsion.

Finally, the T-torsion tensors can be constructed through a
combination of elementary torsion tensors of the forms
\begin{equation}\label{ttor1}
{}^T S_{ab}^{\phan{ab}c}=
V_{[a}e^{A}_{b]}C^{\phantom{A}B}_{A}e^{c}_{B}\,,
\end{equation}
and
\begin{equation}\label{ttor2}
{}^T S_{ab}^{\phan{ab}c}=\epsilon^{ef}_{\phan{ef}ab}
V_{[e}e^{A}_{f]}C^{\phantom{A}B}_{A}e^{c}_{B}\,,
\end{equation}
where $C^{\phantom{A}B}_{A}$ is an arbitrary matrix. By the
null--trace conditions
\begin{equation}\label{condit1}
 V_{[a}e^{A}_{b]}C^{\phantom{A}B}_{A}e^{b}_{B}=0\,,
\end{equation}
and
\begin{equation}\label{condit2}
\epsilon^{ef}_{\phan{ef}ab}
V_{[e}e^{A}_{f]}C^{\phantom{A}B}_{A}e^{b}_{B}=0\,,
\end{equation}
on (\ref{ttor1}) and (\ref{ttor2}),   we obtain 7 constraints on
the matrix $C^{\phantom{A}B}_A$, by fixing $V_a$. As a
consequence,  $C^{\phantom{A}B}_A$ has 9 independent components.
In order to get the 16 components of the T-torsion from the
expressions (\ref{ttor1}) and (\ref{ttor2}), we have to impose a
further condition. If $V^{2}\equiv V^{a}V_{a}\neq 0$, we can
impose that
\begin{equation}\label{condit3}
C^{\phantom{A}B}_{A}e^{A}_{a}e^{b}_{B}V^aV_b=0\,.
\end{equation}
which   reduces one of the constraints following from
(\ref{condit1}) to
\begin{equation}\label{trace}
  C^{\phantom{A}A}_{A}=0\, ,
\end{equation}
If $V$ is a null vector, the constraint (\ref{condit3}) follows
from (\ref{condit1}) and the equation (\ref{trace}) has to  be
imposed as a supplementary constraint on the matrix
$C^{\phantom{A}B}_A$.

~From the previous discussion, it follows that the T-torsion
tensors can be classified according to the nature of the vector
$V_a$ which can be time-like, space-like, or null.
 We label the T-torsion with $Tt$, $Ts$ or $T\ell$ depending on
 whether the $4$-vector $V^a$ is time-like, space-like or light-like.

\section{ Examples}\label{examples}

In a first group of examples, we show how some torsion tensors
 frequently found in literature, can be classified according to the
irreducible tensors classification given above.
\begin{enumerate}
\item Scalar fields $\phi$ produce torsion only in nonminimally
coupled theories with a $\xi \phi^2 R$ term in the Lagrangian
density or in a $R^2$ theory in a {\bf U}$_4$ (where the Ricci
scalar is $R$ coupled to itself). As a result, the torsion is
related to the gradient of the field. For example, in homogeneous
cosmologies, one obtains a $Vt$ tensor. In a Schwarzchild
solution one deals with a $ Vs$ tensor. See for example
\bib{SOL},\bib{german},\bib{cosimo},\bib{desabbata2}.

\item According to  \bib{tsamparlis} and \bib{goenner}, it turns out that
the only torsion tensors compatible with a
Friedmann-Lemaitre-Robertson-Walker universe are of class $ Vt$
and $ At$. A cosmological solution with a torsion tensor of class
$At$ is discussed also in \bib{minkowski}.

\item Examples of torsion of class $V \ell$ and $A\ell$
are found in \bib{baker} to describe null electromagnetic plane
waves.

\item $As$ and $Vs$ tensors introduce anisotropies
in a spacetime, since the spacelike vector yields a privileged
direction.

\item The spin of classical Dirac particles is the source of an
$As$ torsion for massive particles and of an $An$ torsion
for a massless neutrino \bib{hehl}. The $At$ torsion is
generated by tachyon Dirac particles.

\item An example of T-torsion tensors can be found in simple supergravity where torsion is given
in terms of the Rarita--Schwinger spinors (see, for example,
\bib{fvnf}). They contribute also to torsion in the Weyssenhoff spin fluids (see
below and discussion in \bib{bauerle}).

\item The Lanczos tensor was considered in
\bib{baker} as a candidate of torsion in a
non-ECSK theory. It is a sort of Weyl tensor potential and it
bears all the characteristics of a traceless torsion tensor. Then
its properties depend on the symmetries of spacetime.

\item The influence of an $At$ torsion on
cosmological perturbations is discussed in \bib{ZZI}.

\item The helicity flip of fermions can be induced by a $Al$ torsion
\bib{flip}.

\item The same kind of torsion can induce a geometrical contribution
to the Berry phase of Dirac particles \bib{berry}.

\noindent The next group of examples is related to elementary
torsion tensors found in literature.

\item The torsion tensors related to spin, usually found in the
literature, are generated by the Weyssenhoff spinning particle and
the classical Dirac particle. In the first case, the torsion
tensor is a $ Bs$ tensor, in the second case, one has a $ As$
tensor. Spin fluids {\it \`a la} Weyssenhoff can be found in
\bib{hehl},
\bib{deritis},\bib{ray},\bib{obukhov},
\bib{trautman},\bib{tafel},
and have been discussed by  many other authors.

\item Cosmological models with a $Bs$ torsion have been
studied in \bib{tafel}.
\end{enumerate}

\section{ Contributions to
the energy-momentum tensor}\label{emtens}

After straightforward calculation, one obtains that the
contribution of the antisymmetric and vector parts of torsion to
the energy-momentum tensor are respectively proportional to the
following expressions

\beq {}^A {\bf t}_{a}^{b} = 2\sig^b\sig_{a} +
\del_{a}^{b}\sig^c\sig_c \label{emtens1}\,, \eeq
 and
 \beq {}^V
{\bf t}_a^b =  \frac{8}{3}S^b S_a - \frac{4}{3}\del^b_a S^c S_c\,.
\label{emtens2}
 \eeq

The contribution of the T-torsion, when expressed from
(\ref{ttor1}) is

 $$ { }^{T_{1}}{\bf t}^{ab}=-C^{cd}C_{(cd)}V^{a}V^{b} -
V^{c}C_{cd}V^{(a}C^{b)d}
 +\frac{1}{2}V^{c}V_{c}(C_{f}^{\phan{f}a}C^{fb}- C_{\phan{a}f}^{a}C^{bf})
$$

\begin{equation}\label{ttorsionem1}
-\frac{1}{2}V^{c}V^{d}C_{c}^{\phantom{c}a}C_{d}^{\phantom{d}b}
+\frac{1}{2}g^{ab}(C^{cd}C_{(cd)}V^{f}V_{f}-
\frac{1}{2}V^{c}C_{cd}V^{f}C_{f}^{\phantom{f}d})\, ,
\end{equation}
otherwise when the T-torsion is expressed by (\ref{ttor2})

$$ { }^{T_{2}} {\bf t}^{ab}=C^{cd}C_{cd}V^{a}V^{b} +
  V^{d}V_{d} (C_{f}^{\phan{f}a}C^{fb}+ C_{\phan{a}f}^{a}C^{bf})
  -V^{c}V^{d}C_{c}^{\phan{c}a}C_{d}^{\phan{d}b}
$$

\begin{equation}\label{ttorsionem2}
-V^{f}C_{fd}(V^{a}C^{bd} + V^{b}C^{ad}) +
\frac{1}{2}(V^{c}V^{d}C_{cf}C_{\phan{f}d}^{f} -
V^{f}V_{f}C^{cd}C_{cd})\, .
\end{equation}
In (\ref{ttorsionem1}) and in (\ref{ttorsionem2}) we have used
the expression
$C_{a}^{\phantom{a}b}=C_{A}^{\phantom{A}B}e^{A}_{\phantom{A}a}e_{B}^{\phantom{B}b}$.
 The presence of  contributions of distinct irreducible tensors does not
lead to interaction terms except when the two classes of
T-torsion are present.

An elementary torsion tensor,
$S_{ij}^{\phan{\al\be}k}=F_{ij}\Sigma^k$, contributes to
the energy-momentum tensor with a symmetric tensor
proportional to

\beq {}^e{\bf t}_a^b=-2\Sigma^2F^{bc}F_{ac} +F^2\Sigma^b\Sigma_a
-\frac{1}{2} F^2\Sigma^2\del_a^b\,, \eeq

where $\Sigma^2=\Sigma_a\Sigma^a$ and $F^2=F^{ab}F_{ab}$.

Expression (\ref{t12}), through Eqs.(\ref{t4}) and
(\ref{t10}), is the final result involving also ordinary
perfect--fluid matter.

\section{ Contributions of torsion to shear,
expansion, vorticity and acceleration}\label{svea}

It has often been pointed out in the literature how torsion can
modify the behaviour of fluids. In \bib{drls} it was shown that
the presence of a torsion generated by a Weyssenhoff fluid
generalizes the Bernoulli theorem, through an extension of the
definition of the vorticity. In the same way such a modification
of the vorticity has led some authors to argue about the
possibility of having cosmological models with torsion which
could avert the initial singularity \bib{trautman}. An extended
analysis of this problem has been made by restating the
Raychaudhuri equation in the presence of torsion for a
Weyssenhoff fluid \bib{stewart}\bib{esposito}.

In \bib{ddrpss} it was considered an inflationary   Bianchi I
universe in the ECSK theory. In this paper it was shown how
torsion could contribute to an isotropic expansion universe even
in anisotropic universes, if the energy density of spin was
sufficiently large to counterbalance the anisotropic terms. As a
result it followed that this model supplies a rapid isotropization
mechanism of the universe during the inflationary stage.

The previous considerations suggest considering  how the
kinematical quantities are modified by each of the irreducible
components of torsion.

In \bib{Ellis} a gauge invariant and covariant formalism for
cosmological perturbations was formulated. In this derivation an
important r\^ole is attributed to the Raychaudhuri equation. Such
formulation has been extended recently in (\bib{palle}) for the
ECSK theory. It follows that important tests for torsion in the
primordial universe through its effects on the spectrum of
perturbations. A complete study of perturbations for all the
irreducible torsion tensors can be useful to extend this program.

\subsection{\normalsize \bf The kinematical quantities}

One of the consequences of introducing torsion in a
space-time is that the definition of some quantities can be
modified. This is the case of the kinematical quantities,
defined from the following decomposition of  the covariant
derivative of the four velocity $U_a$ \bib{hawking}
\begin{equation}\label{4vel}
 \tilde\nabla_{a}U_{b}= \tilde\sigma_{ab} + \frac{1}{3} h_{ab}
 \tilde\theta+\tilde\omega_{ab} -U_{a}\tilde a_{b}
\end{equation}
where $h_{ab}=g_{ab} +U_a U_b$ and
\begin{equation}\label{expansion}
 \tilde\theta= \tilde\nabla_{a}U^{a}=\theta - 2S^{c}U_{c},
\end{equation}
\begin{equation}\label{shear}
\tilde\sigma_{ab}=h_a^c h_b^d\tilde\nabla_{(c}U_{d)}=\sigma_{ab} +
2h_a^c h_b^dK_{(cd)}^{\phan{(ab)}e}U_{e},
\end{equation}
\begin{equation}\label{vorticity}
 \tilde\omega_{ab}=h_a^c
h_b^d\tilde\nabla_{[c}U_{d]}=\omega_{ab} + 2h_a^c
h_b^dK_{[cd]}^{\phan{[cd]}e}U_{e},
\end{equation}
and the acceleration
\begin{equation}\label{acceleration}
  \tilde a_{c}= U^{a}\tilde\nabla_{a}U_{c}= a_{c}+
  U^{a}K_{ac}^{\phan{ac}d}U_d.
\end{equation}
The quantities without the tilde are those usually defined
in General Relativity.

 We will summarize in Tables I and II the
contributions to this objects given by the irreducible torsion
tensors. \vskip 1truecm

\begin{center}
\begin{tabular}{|c|c|c|c|c|c|c|c|}\hline
&      &      \\
    & ${}^V S_{ab}^{\phan{ab}c}$ & ${}^A
    S_{ab}^{\phan{ab}c}$\\ \hline
    & & \\
  $\tilde\theta=$ &   $\theta-2S^{c}U_{c}$ & $\theta$ \\ \hline
   & &  \\
  $\tilde a_{b}=$ &  $a_{b}-S_{b}-S_{a}U^{a}U_{b}$ & $a_{b}$ \\ \hline
  & &  \\
  $\tilde\omega_{ab}=$ &   $\omega_{ab}$ & $\omega_{ab}-\epsilon_{abcd}\sigma^{d}U^{c}$ \\ \hline
  & & \\
  $\tilde\sigma_{ab}=$ &   $\sigma_{ab}$ & $\sigma_{ab}$ \\ \hline
\end{tabular}
\end{center}

\begin{center}
Table I: {\it Contributions of V-torsion and A-torsion to the
kinematical quantities.}
\end{center}
\

\begin{center}
\begin{tabular}{|c|c|c|c|c|c|c|c|}\hline
&      &
 \\
  & ${}^{T_1} S_{ab}^{\phan{ab}c}$ &${}^{T_{2}} S_{ab}^{\phan{ab}c}$  \\ \hline
    & & \\
  $\tilde\theta=$ & $\theta$ & $\theta$     \\ \hline
   & &  \\
  $\tilde a_{b}=$ &$a_{b}+2V_{[b}C_{a]c}U^{c}U^{a} $& $a_{b}-2\epsilon_{efhb}V^{e}C_{fc}U^{h}U^{a} $       \\ \hline
  & &  \\
  $\tilde\omega_{ab}=$ & $\omega_{ab}+h_{a}^{c}h_{b}^{d}V_{[c} C_{d]}^{\phantom{d]}e}U_{e}$&  $\omega_{ab}+h_{a}^{c}h_{b}^{d}\epsilon_{efcd}V^{e} C^{fg}U_{g}$  \\ \hline
  & & \\
  $\tilde\sigma_{ab}=$ &$\sigma_{ab}- 2 h_{a}^{c}h_{b}^{d}(V_{e}C_{(cd)} - C_{e(c}V_{d)} )U^{e}$&
   $\sigma_{ab}- 2 h_{a}^{c}h_{b}^{d}\epsilon_{efg(a}V^{e}C^{f}_{\phantom{f}d)}U^{g}$   \\ \hline
\end{tabular}
\end{center}
\begin{center}
Table II: {\it Contributions of the two T-torsions to the
kinematical quantities.}
\end{center}

\subsection{\normalsize\bf The Raychaudhuri
equation}

Given the four-velocity $U_a$ ($U_a U^a =-1$), bearing in mind
the identity
\begin{equation}\label{dd}
 U^{b}\tilde\nabla_{c}\tilde\nabla_{b}U_{a}=\tilde\nabla_{c}
 (U^{b}\tilde\nabla_{b}U_{a}) - \tilde\nabla_{c}U^{b}
 \tilde\nabla_{b}U_{a}
\end{equation}
and from Eq.(\ref{riemann})
\begin{equation}\label{ddd}
U^{b}\tilde\nabla_{c}\tilde\nabla_{b}U_{a}=
U^{b}\tilde\nabla_{b}\tilde\nabla_{c}U_{a} +
R_{cba}^{\phan{cba}d}U_{d}U^{b} - 2
U^{b}S_{ab}^{\phan{ab}c} \tilde\nabla_{d}U_{c}
\end{equation}
we find the equation

$$ \frac{1}{3}h_{ca}\tilde\theta +
\tilde\sigma_{ca}+\tilde\omega_{ca}- U_{c}\tilde a_{a}=
\tilde\nabla_{c}\tilde a_{a} $$

$$-\Bigg( \frac{1}{9}h_{ca}\tilde\theta
+\frac{2}{3}\tilde\theta\tilde\sigma_{ca} +
\frac{2}{3}\tilde\theta\tilde\omega_{ca} +
2\tilde\sigma_{c}^{b} \tilde\omega_{ba} $$
$$\tilde\sigma_{c}^{b} \tilde\sigma_{ba} +
\tilde\omega_{c}^{b} \tilde\omega_{ba} -
\frac{1}{3}U_{c}\tilde\theta\tilde a_{a} - U_{c}\tilde
a^{b} \tilde\sigma_{ba}-U_{c}\tilde a^{b}
\tilde\omega_{ba}\Bigg) $$
\begin{equation}\label{penultima}
-R_{cba}^{\phan{cba}d}U_{d}U^{d}-2 U^{b}S_{ab}^{\phan{ab}c}
\left(\frac{1}{3}h_{dc}\tilde\theta +
\tilde\sigma_{dc}+\tilde\omega_{dc}- U_{d}\tilde
a_{c}\right).
\end{equation}
Contracting the indices in Eq.(\ref{penultima}), one obtains
immediately the most general expression for the Raychaudhuri
equation, i.e.
\begin{equation}\label{raychau}
\dot{\tilde\theta}=\tilde\nabla_{c}\tilde a^{c}
-\frac{1}{3}\theta^{2} -\tilde\sigma^{ab}\tilde\sigma_{ab}
+ \tilde\omega^{ab}\tilde\omega_{ab} - R_{ab}U^{a}U^{b} - 2
U^{b}S_{ab}^{\phan{ab}d}
 \left(\frac{1}{3}h_{d}^{a}\tilde\theta +
\tilde\sigma_{d}^{a}+\tilde\omega_{d}^{a}- U_{d}\tilde
a^{a}\right)\, .
\end{equation}
This is the Raychaudhuri equation in its most general form in
presence of torsion. Simpler versions of this equation have
already been discussed in \bib{stewart}, \bib{tafel2},
\bib{esposito} and in
\bib{raych}.

\section{ Conclusions}\label{conclusions}

 As we discussed
in Sec.\ref{examples}there are several ways to build a torsion
tensor. In this paper, we deal with the problem of finding a
geometrical classification of torsion tensors. A decomposition of
torsion into irreducible tensors has been already given (see e.g.
\bib{mccrea},\bib{tsamparlis}, and, for a systematic account,
\bib{affine}). Essentially, one has three classes of tensors:
traceless, vector and totally antisymmetric ones.
 However, we propose to add a classification scheme
to this decomposition. Our proposal is based on the space-time
properties of 4-vectors and bivectors which can be used to
construct these torsion tensors.  According to this
classification, we have shown that it is possible to construct two
tensors of the same irreducible class, with distinct properties,
due to the fact that one can use space-like, time-like, or null
4-vectors.

As a byproduct, we found also a second decomposition and
classification scheme based on elementary torsions. These
elementary tensors are given by the tensor product of simple
bivectors and vectors.  As a consequence, the classification of
these tensors is based on the space-time properties of the simple
bivectors (which we distinguished in {\it electric} and {\it
magnetic} bivectors), and on those of the vectors.

These two classifications, in our opinion, can help to distinguish
the physical situations associated to different torsion tensors.

As an application, we have provided a general scheme for the
modification induced by torsion tensors on kinematical quantities
(such as shear, vorticity, expansion and acceleration). Moreover,
we discussed a general form of the Raychaudhuri equation which can
be physically relevant for the study of many issues such  as the
cosmological perturbations (e.g. see \bib{palle}).

\vspace{3. mm} {\bf Acknowledgements} The authors wish to thank
the referee and Prof. Friedrich Wilhelm Hehl for useful comments
and criticisms which led to a substantial improvement the paper,
 and Dr. Giampiero Esposito for a careful reading of the
manuscript and useful comments.

\vspace{2. cm}

\begin{center}
{\bf References}
\end{center}
\begin{enumerate}
\item\label{hehl}
F.W. Hehl, P. von der Heyde, G.D. Kerlick and J.M. Nester, \rmp
{\bf 48} (1976) 393.
\item\label{GSW}
M.B. Green, J.H. Schwarz, E. Witten, {\it  Superstring Theory}
           (Cambridge University Press, Cambridge, 1987).
 R. Hammond, {\it Il Nuovo Cimento} {\bf 109B} (1994) 319;
           \grg {\bf 28} (1986) 419;
 V. De Sabbata, {\it Ann. der Physik} {\bf 7} (1991) 419.
 V. de Sabbata, {\it Torsion, string tension and quantum gravity},
           in ``Erice 1992'', Proceedings, String Quantum Gravity and Physics
           at the Planck Energy Scale, p. 528.
 Y. Murase, {\it Prog. Theor. Phys.} {\bf 89} (1993) 1331.
 \item\label{GER}
 Yu A. Kubyshin, {\it J. Math. Phys.} {\bf 35} (1994) 310;
 G. German, A. Macias, O. Obregon, \cqg {\bf 10} (1993) 1045;
 C.H. Oh, K. Singh, \cqg {\bf 6} (1989) 1053.
 \item\label{HOW}
P.S. Howe, A. Opfermann, G. Papadopoulos, {\it Twistor
spaces for QKT manifold}, hep-th/9710072; P.S. Howe, G.
Papadopoulos, \pl {\bf 379B} (1996) 80; V. de Sabbata, C.
Sivaram, {\it Il Nuovo Cimento} {\bf 109A}
           1996) 377.
\item\label{LOS}
G. Papadopoulos, P.K. Townsend, {\it Nucl. Phys.} {\bf B444}
           (1995) 245;
C.M. Hull, G. Papadopoulos, P.K. Townsend, \pl {\bf 316B}
           (1993) 291.
\item\label{SIV}
V. De Sabbata, C. Sivaram, {\it Astr. and Space Sci.} {\bf
176} (1991) 141.\\
 H. Goenner and F. M\"uller-Hoissen, \cqg {\bf 1} (1984) 651.\\
P. Minkowski, {\it Phys. Lett.} {\bf 173B} (1986) 247.\\
 R. de Ritis, P. Scudellaro and C. Stornaiolo, \pl {\bf 126A} (1988)
           389.\\
 M.J. Assad, P.S. Letellier, \pl {\bf A145} (1990) 74.\\
 A. Canale, R. de Ritis and C. Tarantino, \pl {\bf 100A} (1984) 178.\\
 A.J. Fennelly and L.L. Smalley, \pl {\bf 129A} (1988) 195.\\
 I.L. Buchbinder, S.D. Odintsov and I.L. Shapiro, \pl {\bf 162B}
           (1985) 92. \\
 C. Wolf, \grg {\bf 27} (1995) 1031.\\
 P. Chatterjee, B. Bhattacharya, {\it Mod. Phys. Lett.} {\bf A8}
           (1993) 2249.
\item\label{MAR}
A. Dobato and A. L. Maroto, {\it Leptogenesis from torsion
through the axial anomaly}, hep-ph/9705434.
\item\label{FIG}
B.D.B. Figueiredo, I. Damiao Soares and J. Tiomno, \cqg
{\bf 9}(1992) 1593; L.C. Garcia de Andrade, {\it Mod. Phys.
Lett.} {\bf 12A} (1997) 2005; O. Chandia, \pr {\bf D55}
(1997) 7580; P.S. Letelier, \cqg {\bf 12} (1995) 471; S.
Patricio, S. Letelier, \cqg {\bf 12} (1995) 2221; R.W.
K\"uhne, {\it Mod. Phys. Lett.} {\bf A12} (1997) 2473; D.K.
Ross, {\it Int. J. Theor. Phys.} {\bf 28} (1989) 1333.
\item\label{baker}
W. M. Baker, W. K. Atkins, W. R. Davis, \ncim {\bf B 44}
(1978) 1; W. M. Baker, W. K. Atkins, W. R. Davis, \ncim
{\bf B 44} (1978) 17; W. M. Baker, W. K. Atkins, W. R.
Davis, \ncim {\bf B 44} (1978) 23.
 \item\label{OTM}
O. Moritsch, M. Schweda, and S.P. Sorella, \cqg {\bf 11} (1994)
1225.
\item\label{schouten} J.A. Schouten {\it Ricci Calculus}
(Springer-Verlag, Berlin 1954)
\item\label{mtw} C.W. Misner, K.S. Thorne and J.A. Wheeler,
{\it Gravitation}  (Freeman, San Francisco, 1973).
\item\label{mccrea}
J. Dermott McCrea, \cqg {\bf 9} (1992) 553.
\item\label{tsamparlis}
M. Tsamparlis \pr {\bf D 24} (1981) 1451; M. Tsamparlis \pl {\bf A
75} (1979) 27.
\item\label{affine} F.W. Hehl, J.D. McCrea, E.W. Mielke, Y.
Ne'eman, Phys. Rep. {\bf 258} (1995), 1.
\item\label{goenner}
H. Goenner and F. M\"uller-Hoissen, \cqg {\bf 1} (1984) 651.
\item\label{BAE}
P. Baekler, E.W. Mielke, F.W. Hehl, {\it Il Nuovo Cimento}
           {\bf 107B} (1992) 91.
\item\label{trautman}
A. Trautman {\it Nature (Phys. Sci.)}  {\bf 242} (1973) 7.
\item\label{deritis}
R. de Ritis, M. Lavorgna, G. Platania, C. Stornaiolo \pr {\bf D
28} (1983) 713;   R. de Ritis, M. Lavorgna, G. Platania, C.
Stornaiolo \pr {\bf D 31} (1985) 1854.
\item\label{drls}R. de Ritis, M. Lavorgna, C. Stornaiolo Phys.Lett. {\bf
95A}(1983) 425.
\item\label{ray}
J.R Ray and L. Smalley \pr {\bf D 27} (1983) 1383;  J.R Ray
and L. Smalley \prl {\bf 49} (1982) 1059.
\item\label{obukhov} Yu. N. Obukhov and V.A, Korotky \cqg {\bf 4}
(1987), 1633.
\item\label{hawking}
S.W. Hawking and G.F.R. Ellis, {\it The large scale structure of space-time},
Cambridge Univ. Press 1973, Cambridge.
\item\label{desabbata2}
V. de Sabbata and C. Sivaram, \ass {\bf 165} (1990) 51; V.
de Sabbata and C. Sivaram, \ass {\bf 176} (1991) 141;
 V. de Sabbata, \ncim {\bf A 107} (1994) 363.
\item\label{minkowski}
P. Minkowski, \pl {\bf  B 173} (1986) 247.
\item\label{SOL}
 H. Soleng, {\it Class. Quan. Grav.} {\bf 5} (1988) 1489.
\item\label{german} G. German, \pr {\bf D32} (1985) 3307.
\item\label{cosimo}
C. Stornaiolo, {\it La cosmologia nella teoria ECSK}, PhD
thesis, Universit\`a di Napoli ``Federico II" (1987) Napoli.
\item\label{bauerle} G.A B\"auerle and C.J. Haneveld, {\it
Physica} {\bf 121A} (1983) 121.
\item\label{fvnf} D.Z. Freedman, P. van Nieuwenhuizen, S. Ferrara, \pr {\bf D13}
(1976) 3214; D.Z. Freedman, P. van Nieuwenhuizen, \pr {\bf D14}
(1976) 912.
\item\label{ZZI}
S. Capozziello and C. Stornaiolo,  {\it Il
Nuovo Cimento} {\bf B} {\bf 113} (1998) 879.
\item\label{flip}
S. Capozziello, G. Iovane, G. Lambiase, and C. Stornaiolo,
{\it Europhys. Lett.} {\bf 46} (1999) 710.\\
R.T. Hammond, \cqg {\bf 13}  (1996) 1691.
\item\label{berry}
S. Capozziello,  G. Lambiase, and C. Stornaiolo,
  {\it Europhys. Lett.} {\bf 48}  (1999) 482.
\item\label{tafel}
J. Tafel, {\it  Acta Physica Polonica} {\bf B6} (1975) 537.
\item\label{stewart}
J. Stewart and P. Hajicek {\it Nature (Phys. Sci.)}{\bf 244}
(1973) 96.
\item\label{tafel2}
J. Tafel Phys. Lett. {\bf 45A} (1973) 341.
\item\label{raych} A.K. Raychaudhuri ``Theoretical
Cosmology'' (1979) Clarendon Press, Oxford U.K.
\item\label{esposito} G. Esposito, Fortschritte der Physik {\bf
40} (1992) 1.

\item\label{ddrpss} M.~Demianski, R.~De Ritis,G.~Platania, P.~Scudellaro and C.~Stornaiolo,
Phys.\ Rev.\  {\bf D35} (1987) 1181.
\item\label{Ellis}G.~F.~Ellis and M.~Bruni, Phys.\ Rev.\  {\bf D40} (1989)
1804;  G.~F.~Ellis, J.~Hwang and M.~Bruni, Phys.\ Rev.\ {\bf
D40}, 1819 (1989); G.~F.~Ellis, M.~Bruni and J.~Hwang, Phys.\
Rev.\  {\bf D42} (1990) 1035.
\item\label{palle} D. Palle Nuovo Cim. {\bf B114}  (1999) 853.

\end{enumerate}
\vfill

\end{document}